\documentclass[aps,prl,twocolumn,superscriptaddress,showpacs]{revtex4}

\usepackage[dvips]{graphicx}
\usepackage{latexsym}

\begin{document}

%Title of paper

\title{ First-principles-based $\pm s$-wave modelling for iron-based
superconductors:
Studies for specific heat and nuclear magnetic relaxation rate }

\author{
N. Nakai
}
\email[]{nakai.noriyuki@jaea.go.jp}
\affiliation{
CREST(JST), 4-1-8 Honcho, Kawaguchi, Saitama 332-0012, Japan
}
\affiliation{
CCSE, Japan Atomic Energy Agency, 6-9-3 Higashi-Ueno, Taito-ku, Tokyo 110-0015, Japan
}
\author{
H. Nakamura
}
\affiliation{
CREST(JST), 4-1-8 Honcho, Kawaguchi, Saitama 332-0012, Japan
}
\affiliation{
CCSE, Japan Atomic Energy Agency, 6-9-3 Higashi-Ueno, Taito-ku, Tokyo 110-0015, Japan
}
\affiliation{
TRIP(JST), 5 Sanban-cho, Chiyoda-ku, Tokyo 110-0075, Japan
}
\author{
Y. Ota
}
\affiliation{
CREST(JST), 4-1-8 Honcho, Kawaguchi, Saitama 332-0012, Japan
}
\affiliation{
CCSE, Japan Atomic Energy Agency, 6-9-3 Higashi-Ueno, Taito-ku, Tokyo 110-0015, Japan
}
\author{
Y. Nagai
}
\affiliation{
TRIP(JST), 5 Sanban-cho, Chiyoda-ku, Tokyo 110-0075, Japan
}
\affiliation{
Department of Physics, University of Tokyo, Tokyo 113-0033, Japan
}
\author{
N. Hayashi
}
\affiliation{
CREST(JST), 4-1-8 Honcho, Kawaguchi, Saitama 332-0012, Japan
}
\affiliation{
N2RC, Osaka Prefecture University, 1-2 Gakuen-cho, Naka-ku, Sakai 599-8570, Japan
}
\author{M. Machida
}
\affiliation{
CREST(JST), 4-1-8 Honcho, Kawaguchi, Saitama 332-0012, Japan
}
\affiliation{
CCSE, Japan Atomic Energy Agency, 6-9-3 Higashi-Ueno, Taito-ku, Tokyo 110-0015, Japan
}
\affiliation{
TRIP(JST), 5 Sanban-cho, Chiyoda-ku, Tokyo 110-0075, Japan
}

\date{\today}

\begin{abstract}

In order to consistently explain controversial experimental results 
on superconducting states observed by different probes in typical iron-based superconductors, 
we construct a realistic multi-band $\pm s$-wave pairing model 
by combining the quasiclassical formalism with the first-principles calculation.
The model successfully resolves the controversies 
in contrast to the fact that simplified models 
such as two-band $\pm s$-wave one fail to do.
A key in the model is the existence of relatively small gaps 
which leads to material-dependent peculiarities.   

\end{abstract}

\pacs{74.20.-z, 74.70.Dd, 74.25.Jb
% 74.20.-z Theories and models of superconducting state 
% 74.25.Jb Electronic structure 
% 74.70.Dd Ternary, quaternary, 
% and multinary compounds (including Chevrel phases, borocarbides, etc.)  
}

\maketitle

The discovery of the iron-pnictide superconductor LaFeAs(O$_{1-x}$F$_x$) \cite{kamihara} made 
a striking impact on materials science, because this compound includes the element of the most 
familiar ferromagnetic metal, Fe, as a main component. 
The transition temperature $T_{\rm c}$ in the so-called ``1111'' compounds $R$FeAs(O$_{1-x}$F$_x$) ($R$=Pr, Nd, Sm)
exceeds 50K, which is the highest except for high $T_{\rm c}$ cuprates.
In addition to the high transition temperature, the variety of related materials is quite rich.
For example, ``122'' compounds ($A_{1-x}B_{x}$)Fe$_2$As$_2$ ($A$=Ba, Sr, Ca, $B$=K, Cs, Na) and 
``11'' compounds Fe(Se$_x$Te$_{1-x}$) are the typical family materials, 
whose element substitutions are widely possible \cite{rotter,ishida}. 
In particular, the superconductivity is surprisingly robust against substitutions of Ni and Co for Fe.

In contrast to the discovery rush of family compounds, 
their superconducting states still remain elusive. 
There is no established pairing-symmetry model explaining all experimental results consistently.
In the early days, puzzling experimental data were reported. 
In spite of sharp resistivity drops and 
clear Meissner signals at the superconducting transition $T_{\rm c}$, 
the jump of $C/T$, where $C$ is the specific heat 
and $T$ is the temperature, was hardly observable in 1111 compounds.
The reason was initially ascribed to a large phonon contribution which masks an electronic one. 
Afterwards, a small jump at $T_{\rm c}$ and the concave-down  temperature dependence of $C/T$ below $T_{\rm c}$ 
were confirmed in not only 1111 compounds \cite{ding,mu} but also structurally equivalent 
LaFeP(O$_{1-x}$F$_x$) \cite{kohama}. 
This fact strongly suggests the existence of rather small superconducting gap \cite{evt} 
in addition to large main gaps in the cases of $s$-wave gap. 
In this paper, we clarify that such a multi-gap structure consistently explains 
all experimental observations of 1111 compounds by means of a realistic five-band 
$\pm s$-wave pairing model based on a first-principles calculation. 
A striking result of the model is a natural reproduction 
of the nuclear magnetic relaxation rate $1/T_1$
below $T_{\rm c}$.

On the other hand, 122 compounds experimentally exhibit large jumps and exponential behavior 
in $C/T$ like conventional superconductors \cite{mu,budko,kurita,ronning}. 
Moreover, the power law in the $T$-dependence of $1/T_1$ below $T_{\rm c}$ is 
different from that of 1111 compounds \cite{ishida,fukazawa1,yashima,kobayashi}.
In fact, an angle resolved photoemission spectroscopy (ARPES) study
reported that all gaps fully open and the difference between their gap amplitudes
is not so significant \cite{nakayama}. 
In this case, we reveal that the multi-gap structure according to ARPES data  
can consistently explain the specific heat and 
$1/T_1$ data without any assumptions except 
for $\pm s$-wave paring symmetry \cite{mazin,kuroki}. 
From the present analyses on 1111 and 122 compounds, it is 
found that the existence of the relatively small gap 
gives rise to the material variety.
We believe that this fact has an key role on 
the quest for the superconducting mechanism.

Let us present a procedure to construct the realistic multi-band model.
For the band structure around the Fermi level $E_{\rm F}$, 
we perform a first-principles calculation \cite{vasp}, 
which provides multiple Fermi-surfaces 
and their density of states (DOS) at $E_{\rm F}$ depending on the target materials.
When evaluating superconducting gaps for multi-band superconductors,
we examine all the experimental data and select key data.
ARPES measurements are successful for 122 compounds \cite{nakayama}.
Therefore, we can directly fit ARPES data to determine the gap amplitudes on each band
in the 122 case.
On the other hand, ARPES \cite{kondo} measurements 
are technically difficult for 1111 compounds.
Instead, 
to estimate the gap amplitudes
we adopt data of $C/T$ \cite{kohama} 
and the penetration depth \cite{hashimoto1,martin1} in the 1111 case. 
In fact, those data clearly suggest that the pairing symmetry is full-gap but a small 
single or small multi gaps coexist with main large gaps \cite{hashimoto1,martin1,martin2,ding2,luo,tanatar}.

\begin{figure*}[tb]
\includegraphics[width=15cm,keepaspectratio]{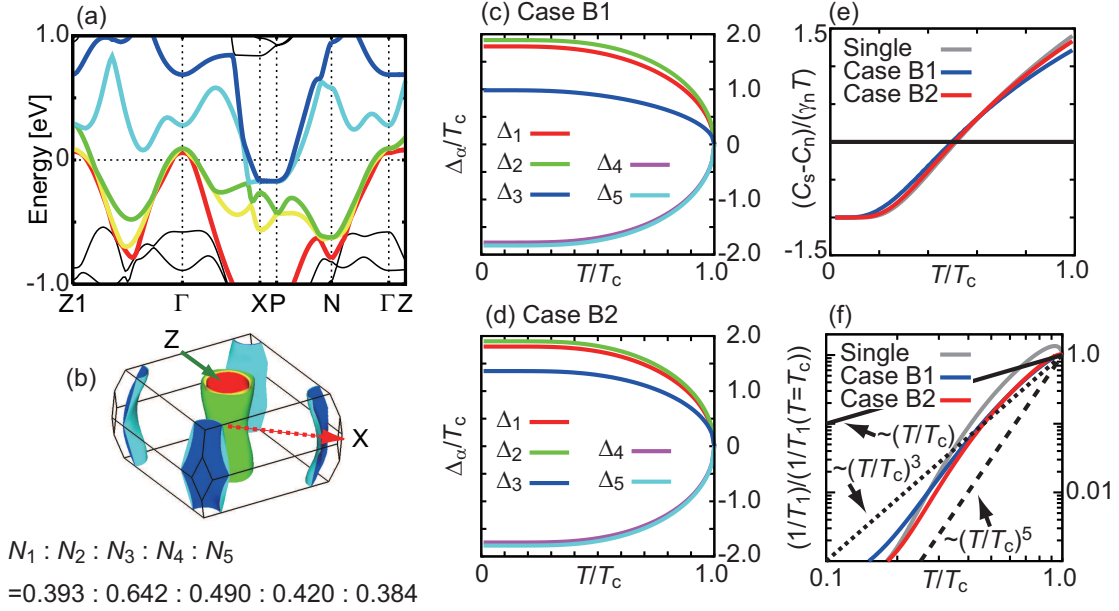}
\caption{
(color online)
(a) The band structure calculated by the generalized gradient approximation 
using structural measurement values of BaFe$_2$As$_2$.
(b) The Fermi surfaces and the density of states at the Fermi energy.
Indices ($\alpha$=1, 2, 3, 4, 5) are assigned from $\Gamma$(zone center) to X.
Temperature dependences of the superconducting pair-potential $\Delta_\alpha$ 
are displayed in (c) and (d).
Here,  ``B'' of Case B1 or B2 stands for BaFe$_2$As$_2$.
(e) Temperature dependences of $(C_{\rm s}-C_{\rm n})/T$. $C_{\rm s(n)}$ is the specific heat of the superconducting state (normal state). 
(f) Temperature dependences of the nuclear magnetic relaxation rate $1/T_1$.
}
\label{fig1}
\end{figure*}
\begin{figure*}[tb]
\includegraphics[width=15cm,keepaspectratio]{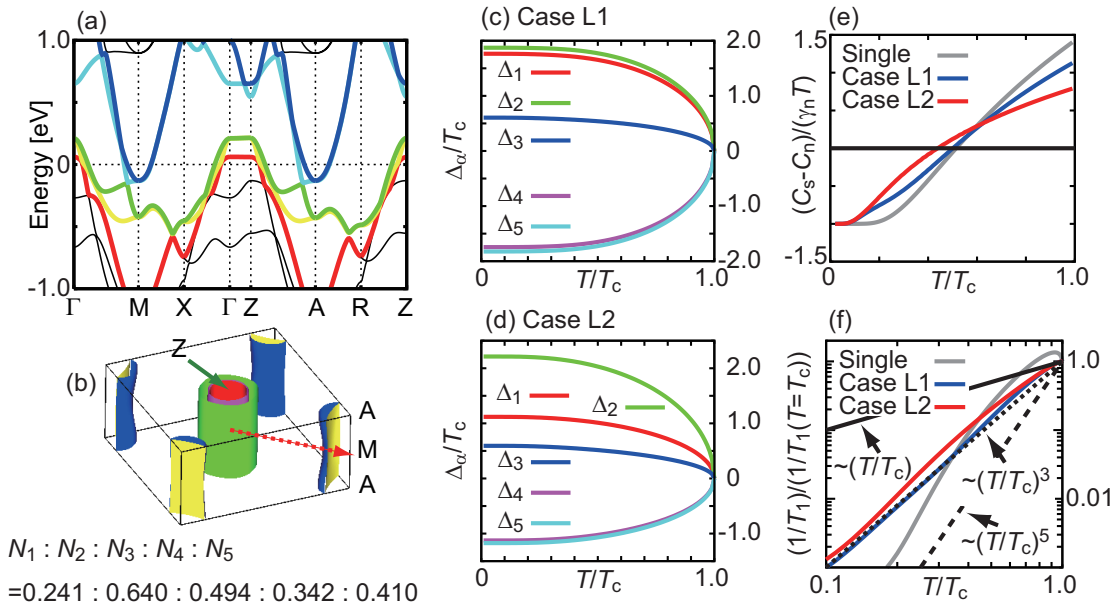}
\caption{
(color online)
(a) The band structure calculated by the generalized gradient approximation 
using structural measurement values of LaFeAsO.
(b) The Fermi surfaces and the density of states at the Fermi energy.
Indices ($\alpha$=1, 2, 3, 4, 5) are assigned from $\Gamma$(zone center) to M.
Temperature dependence of the superconducting pair-potential $\Delta_\alpha$ in (c) and (d).
Here,  ``L'' of Case L1 or L2 stands for LaFeAsO.
(e) Temperature dependences of $(C_{\rm s}-C_{\rm n})/T$. 
$C_{\rm s(n)}$ is the specific heat of the superconducting state (normal state). 
(f) Temperature dependences of the nuclear magnetic relaxation rate $1/T_1$.
}
\label{fig2}
\end{figure*}
We describe the expression for $C/T$ here, 
while we refer readers to Refs.~\cite{mitrovic, bang, nagai}
for $1/T_1$.
The $T$-dependence of $C/T$ is calculated by the second derivative of the free energy.
The free energy can be evaluated by the quasiclassical theory of superconductivity \cite{eilenberger,kopnin},
which is a mean field treatment convenient in evaluating superconducting properties. 
To calculate $C/T$ and $1/T_1$,
we need the $T$-dependence of the multiple superconducting pair-potential $\Delta_{\alpha}$ on each band.
For this purpose, 
we solve the gap equations
for multi-band superconductors \cite{suhl}. 
With Matsubara frequency 
$\mbox{$\omega_n=(2n+1)\pi T$}$, 
the gap equations are written as
\begin{eqnarray}
\Delta_\alpha
&=&
2\pi T\sum_{\omega_n>0}
\sum_{\beta}
\lambda_{\alpha\beta}
f_{\beta}(i\omega_n),
\\
\Delta_\alpha^\ast
&=&
2\pi T\sum_{\omega_n>0}
\sum_{\beta}
\lambda_{\alpha\beta}
f_{\beta}^\dagger(i\omega_n).
\end{eqnarray}
where $\alpha$ and $\beta$ stand for the band index, and 
$\lambda_{\alpha\beta}$ is 
the effective coupling constant.
In addition, $\mbox{$\lambda_{\alpha\beta}=N_\beta \lambda_{\beta\alpha} /N_\alpha$}$
with $N_\alpha$ being DOS at $E_{\rm F}$ for $\alpha$-band.
$\lambda_{\alpha\alpha}$ comes from the intra-band interaction, 
and  $\lambda_{\alpha\beta}$, 
where $\mbox{$\alpha\ne\beta$}$, 
gives the pair-hopping between the different bands.
The effective coupling constants $\lambda_{\alpha\beta}$'s are employed 
as parameters to reproduce the experimental results for the superconducting gaps. 
Again, we note the selected experiment type depending on a kind of the compounds, 
i.e., ARPES \cite{nakayama} for the 122 compounds, 
and the specific heat \cite{kohama} 
and the penetration depth \cite{hashimoto1,martin1} for the 1111 compounds.

The quasiclassical Green's functions 
$g_{\alpha}(i\omega_n)$,
$f_{\alpha}(i\omega_n)$, and
$f_{\alpha}^\dagger(i\omega_n)$
follow the Eilenberger equations as,
\begin{eqnarray}
\omega_n f_{\alpha}(i\omega_n)
&=&
\Delta_\alpha
g_{\alpha}(i\omega_n),\\
\omega_n f_{\alpha}^\dagger(i\omega_n)
&=&
\Delta_\alpha^\ast
g_{\alpha}(i\omega_n),\\
g^2_{\alpha}(i\omega_n)
&=&
1-f_{\alpha}(i\omega_n)f_{\alpha}^\dagger(i\omega_n),
\end{eqnarray}
where Re $g_{\alpha}(i\omega_n)>0$ for $\omega_n >0$.
The free energy difference between the superconducting and normal states
\cite{eilenberger,kopnin,watanabe,kogan}, 
$F_{sn}=F_{\rm super}-F_{\rm normal}$, is expressed as 
\begin{equation}
F_{sn}=
-2\pi T
\sum_\alpha
\sum_{\omega_n>0}
N_\alpha
\bigg[
\frac{1-g_\alpha(i\omega_n)}{1+g_\alpha(i\omega_n)}
\Delta_\alpha^\ast
f_{\alpha}(i\omega_n)
\bigg],
\end{equation}
which demands the solutions of Eqs.\ (3), (4), and (5),
\begin{eqnarray}
f_{\alpha}(i\omega_n)
&=&
\frac{\Delta_\alpha}{\sqrt{\omega_n^2+|\Delta_\alpha|^2}},
\\
f_{\alpha}^\dagger(i\omega_n)
&=&
\frac{\Delta^\ast_\alpha}{\sqrt{\omega_n^2+|\Delta_\alpha|^2}},
\\
g_{\alpha}(i\omega_n)
&=&
\frac{\omega_n}{\sqrt{\omega_n^2+|\Delta_\alpha|^2}}.
\end{eqnarray}
Note that $\mbox{$f^{(\dagger)}_\alpha(i\omega_n)\to 0$}$ and 
$\mbox{$g_\alpha(i\omega_n)\to 1$}$
when $\mbox{$\omega_n \gg |\Delta_\alpha|$}$.
The cut-off frequency 
$\omega_c$ is introduced as $\sum_{\omega_n>0}^{\omega_c}$
in Eqs. (1), (2) and (6)  \cite{detail}.
Equation (6) indicates that 
$F_{sn}$ can be directly evaluated by DOS $N_\alpha$ obtained 
from first-principle calculations and the gap values $\Delta_\alpha$.

Since $C/T$ at constant volume is generally obtained by 
$\mbox{$
{C}/{T}=
{\partial S}/{\partial T}
=
-{\partial^2 F}/{\partial T^2}
$}$, 
the specific heat in the superconducting state is expressed as 
\begin{eqnarray}
\frac{C_s-C_n}{T}=
-
\frac{\partial^2 F_{sn}}{\partial T^2},
\end{eqnarray}
where $C_{\rm s(n)}$ is the specific heat of the superconducting state (normal state).
We can rewrite $C_{\rm n}/T$ as $\gamma_{\rm n}$, 
which is the Sommerfeld coefficient.
Thus, we numerically calculate ${\partial^2 F_{sn}}/{\partial T^2}$ to obtain $C_{\rm s}/T$.
As for $1/T_{1}$ \cite{mitrovic,bang,nagai}, 
we calculate it using $N_\alpha$ and $\Delta_\alpha$ obtained here.

Let us present the calculated results. The first focus is 122 compounds.
Figure~1(a) shows the band structure of BaFe$_2$As$_2$.
This result is obtained by using the generalized gradient approximation (GGA)
based on the measured structural data \cite{rotter}.
The obtained Fermi surfaces are displayed in Fig.~1(b).
The calculation gives DOS's at $E_{\rm F}$, which 
are input parameters in the gap equations (1) and (2) and 
the free energy (6).
Each DOS at $E_{\rm F}$ is written as $N_\alpha$, 
where $\alpha$ is numbered as $\alpha=1, 2, 3, 4, 5$
from $\Gamma$ point (zone center) to ${\rm X}$ as shown in Fig.~1(b), and 
$\alpha=1$ to $3$ ($4$ to $5$) correspond to hole (electron) bands.
Throughout this paper, we adopt $\pm s$-wave pairing model since 
other choices fail to reproduce the experimental data consistently.
The positive (negative) sign is assigned to $\Delta_\alpha$ of hole (electron) bands 
($\Delta_\alpha$ are assumed to be real).
The $T$-dependence of $\Delta_\alpha$ is calculated so as to follow the ARPES result \cite{nakayama} (Case B1)
as shown in Fig.~1(c). 
On the other hand, in Fig.~1(d) we show, for comparison, the result (Case B2) with the minimum $|\Delta_{3}|$ being slightly bigger than the ARPES result.
$\mbox{Figure 1(e)}$ shows the $T$-dependences of $C/T$.
Both Cases B1 and B2 show no significant difference 
from the weak coupling single-band Bardeen-Cooper-Schrieffer (BCS) result.
One of the reasons is that the weighting of the small-gap band is small compared to total one, 
i.e., the ratio 
$N_3/N_t$ is 0.2, where $N_t$ is the total DOS 
$\mbox{($N_t=\sum_\alpha N_\alpha$)}$.
Moreover, the gap-amplitude difference between the minimum $|\Delta_3|$ and the maximum $|\Delta_2|$
is not so large, where $\mbox{$|\Delta_3|/|\Delta_2| \sim 0.5$ $(\sim 0.7)$}$ for Case B1 (B2).
On the other hand, 
the influence of $|\Delta_2|/T_{\rm c}\sim 2$, which is bigger than the single-band's value $1.76$,
slightly enhances $C/T$ at $T_{\rm c}$. 
As a result, the difference in $C/T$ becomes small among Case B1, B2 and single-band case.
The $T$-dependences of $C/T$ in Cases B1 and B2 are consistent with the experimental observations \cite{mu}.

On $T$-dependence of $1/T_1$, both the cases show significant differences 
from the single-band BCS case as shown in Fig.~1(f).
The coherence peak just below $T_{\rm c}$ is absent in both Cases B1 and B2, because the cancellation 
between ``$+$'' and ``$-$'' signs of $\Delta_{\alpha}$ is effective.
In contrast, in the single-band BCS case, even if the damping rate of the quasiparticle 
is taken large as $\eta=0.1T_{\rm c}$, the peak is clearly identified as seen in Fig.~1(f).
Moreover, the low-lying excitation arising from the small gap $|\Delta_3|$ alters $T$-dependence of $1/T_1$
compared to the single-band $s$-wave case. 
We point out that the five-band model Case B1 successfully reproduces 
the experimental results of $C/T$ and $1/T_1$ \cite{mu,yashima}.

We next turn to 1111 compounds.
At first, based on GGA with measured structural parameters \cite{cruz}, 
we obtain the band structure of 
LaFeAsO as shown in Fig.~2(a).
The band index is also numbered as $\alpha=1, 2, 3, 4, 5$
from $\Gamma$ (zone center) to ${\rm M}$.
The five Fermi surfaces are displayed in Fig.~2(b), where
$N_1$, $N_2$, $N_3$ ($N_4$, $N_5$) are the DOS at 
$E_{\rm F}$ of hole (electron) bands.
We prepare two types of multi-gap structures, Cases L1 and L2, whose
$T$-dependences of $\Delta_\alpha$ are plotted in Figs.\ 2(c) and 2(d), respectively.
These are estimated by experimental data of $C/T$ \cite{kohama,mu,ding} and the penetration depth \cite{martin1},
because direct experimental ARPES data of the gaps are not presently available in 1111 compounds.
Case L1 gives four large gaps and one small gap, while 
Case L2 considers a medium gap between the maximum and minimum gap-amplitudes.
Figure~2(e) shows $T$-dependence of $C/T$ for both Cases L1 and L2.
When the amplitude difference between the minimum and maximum gaps is large,
the jump of $C/T$ at $T_{\rm c}$ decreases.
For comparison, we note that the small gap in the 122 compounds   
as shown in Fig.~1(e) is not enough to reduce the jump of $C/T$ 
like the present 1111 compound case.
In addition, Case L2 reproduces the concave-down behavior as observed in
LaFeP(O$_{1-x}$F$_x$) \cite{kohama}.
Thus, the small-jump feature and the concave-down behavior in $C/T$ 
suggest not only the existence of a small-gap but also that of a medium-gap band, 
whose contributions are significant compared to 122 compounds.
We then expect that Case L2 is the most possible candidate. 

Figure~2(f) shows $T$-dependence of $1/T_1$ for both Cases L1 and L2.
We find that both Cases L1 and L2 
surprisingly exhibit $T^3$-behavior of $1/T_1$ up to the experimentally accessible low $T$.
The low-lying excitations due to the small gap push
the exponential behavior of $1/T_1$ into a further lower temperature region.
Moreover, the peak just below $T_{\rm c}$ does not appear 
because of the cancellation due to $\pm$ signs \cite{nagai}.
The most of 1111 compounds show $T^3$ dependence in $1/T_1$ below $T_{\rm c}$ 
\cite{ishida, yashima},
which suggests that Case L2 is the best as noted in Fig.~2(f).

In conclusion, we examined the validity of $\pm s$-wave scenario for typical 
iron-based superconductors (122 and 1111 compounds) 
through the realistic model using the quasiclassical formalism combined with 
first-principles calculations.
Consequently, we found that any anomalous properties observed in the specific heat and 
the nuclear magnetic relaxation rate are reproducible 
without any extrinsic assumptions, 
i.e., all required is the properly-evaluated gap amplitude of each band.

Finally, we add a note that
the momentum dependence of the gap amplitudes is reported by 
some recent experiments \cite{yamashita,fukazawa2,hashimoto2}
of the related compound.
Such an anisotropy can be easily implemented in the present framework.

\begin{acknowledgments}
We acknowledge the fruitful discussions 
with H.~Fukazawa.
Y.N.\ is supported 
by Grant-in-Aid for JSPS Fellows,
and M.M.\ is supported by JPSJ Core-to-Core
Program-Strategic Research Networks, 
``Nanoscience and Engineering in Superconductivity (NES)''.

\end{acknowledgments}
\bibliography{basename of .bib file}

\end{document}